\documentclass[twocolumn,aps,prl]{revtex4}
\usepackage{graphicx}
\begin{document}

\preprint{This line only printed with preprint option}

\title{On the Bose-Einstein Condensation of Magnons in Cs$_2$CuCl$_4$}

\author{S. E. Sebastian$^1$, V. S. Zapf$^2$, N. Harrison$^2$, C. D. Batista$^3$, P. A. Sharma$^2$, M. Jaime$^2$, I. R. Fisher$^1$, A. Lacerda$^2$}

\affiliation{$^1$Geballe Laboratory for Advanced Materials and
Department of Applied Physics, Stanford University, Stanford, CA
94305}

\affiliation{$^2$MST-NHMFL, Los Alamos National Laboratory, Los
Alamos, NM 87545}

\affiliation{$^3$Theoretical Division, Los Alamos National
Laboratory, Los Alamos, NM 87545}

\date{\today}

\maketitle

In a recent paper \cite{Radu}, Radu \textit{et al.} report
experimental results they claim to support Bose-Einstein
condensation (BEC) of magnons in Cs$_2$CuCl$_4$. It is true that an
experimentally measured critical power law scaling exponent in
agreement with the BEC universality class would support the
realization of a BEC in magnetic systems that order as a canted
antiferromagnet. It can be shown, however, that the claim of Radu
{\it et al.} is overstated in this instance, because their
determination of the critical exponent $\phi$ relies on  a
model-dependent theoretical approximation to the critical field
$H_{\textrm{c1}}$ for which the associated errors are neglected. We
show that when these errors are included, the uncertainty in the
obtained exponent is so large that the available experimental data
cannot be used to differentiate between contending universality
classes.

A two parameter fit to only a few data points delineating the
critical ordering temperature ($T_\textrm{c}$) versus magnetic field
($H$)in the vicinity of the quantum critical point (QCP), to the
power law
\begin {equation}
T_\textrm{c}~\sim~(H~-~H_{\textrm{c1}})^{\frac{1}{\phi}} \label
{powerlaw}
\end {equation}
with both $H_{\textrm{c1}}$ and the critical exponent $\phi$ varying
has been shown to be unreliable \cite{Nohadani, Sebastian}. An
independent experimental determination of $H_{\textrm{c1}}$ is
therefore required to obtain an accurate estimate of $\phi$. Given
that neutron scattering measurements on Cs$_2$CuCl$_4$ presented in
Ref.~\cite{Coldea} have provided such a determination, yielding
$H_{\textrm{c1}}=$~8.44~$\pm$~0.01~T, this would be an appropriate
value to use in the fit to Eqn.~(\ref{powerlaw}). Radu {\it et al.}
instead use a value of $`H_{\textrm{c1}}$'=~8.51~T in their fit to
Eqn.~(\ref{powerlaw}), calculated using an approximate theoretical
Hamiltonian, that is subsequently assumed to have {\it zero error}
in their analysis. This assumption has two principal inaccuracies.
The first is that the model Hamiltonian neglects higher order
interactions, thereby introducing an unknown systematic error in
$`H_{\textrm{c1}}$'. The second is that the exchange couplings used
in its computation have significant experimental uncertainty,
introducing a large error in $`H_{\textrm{c1}}$'. We obtain
$`H_{\textrm{c1}}$'=~8.51~$\pm$~0.12~T on using the published errors
in the exchange interactions~\cite{Coldea}.

\begin{figure}[htbp]
\includegraphics[width=0.38\textwidth]{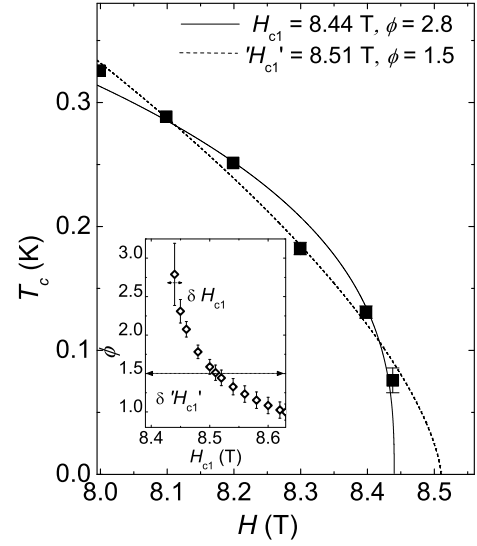}
\caption{Points on the ordering phase boundary from the experimental
data in \cite{Radu}. The solid line represents the best fit to
$\phi$ using the experimentally measured value of
$H_{\textrm{c1}}=$~8.44~T from \cite{Coldea}. The dashed line
represents the best fit to $\phi$ using the theoretical estimate of
$`H_{\textrm{c1}}$'=~8.51~T as per the analysis technique used in
\cite{Radu}. The inset shows the variation in the fit value of
$\phi$ with the value of $H_{\textrm{c1}}$.} \label{fig1}
\end{figure}

Fig.~\ref{fig1} shows fits of Eqn.~(\ref{powerlaw}) to the
experimentally measured phase boundary data points using both the
experimental value of $H_{\textrm{c1}}=$~8.44~$\pm$~0.01~T of Coldea
{\it et al.}~\cite{Coldea} and the theoretical estimate of
$`H_{\textrm{c1}}$'=~8.51~$\pm$~0.12~T, yielding
$\phi=$~2.8~$\pm$~0.4 and $\phi=$~1.5~$\pm$~0.9 respectively, on
considering the dominant contribution to the error:
$\delta\phi~=~\frac{d\phi}{dH}\mid_{_{H_{c1}}}\delta
H_{\textrm{c1}}$. The single most important factor responsible for
the very large error of $\sim~60\%$ in the case of the latter as
compared to the error of $\sim~14\%$ in the former fit, is the
extreme sensitivity of the fit $\phi$ to the theoretical estimate of
the critical field $`H_{\rm c1}$', as depicted graphically in the
inset to Fig.~\ref{fig1}.

Given the substantial uncertainty in the value of $\phi$ that is
obtained from a rigorous analysis, it is clear that the available
experimental data do not favor the 3$d$ BEC universality class
($\phi$~=~1.5) over other possibilities, including the 3$d$ Ising
universality class ($\phi$~=~2).

\end{document}